\documentclass[conference]{IEEEtran}
\IEEEoverridecommandlockouts

\usepackage{amsmath,amssymb,amsfonts}

\usepackage{graphicx}
\usepackage{textcomp}
\usepackage{xcolor}

\usepackage{cite,citesort}
\usepackage{makecell}
\usepackage{hhline}

\usepackage{floatpag}
\usepackage{float}

\usepackage{algorithm}
\usepackage{algorithmicx}
\newcommand{\Input}{\item[\textbf{Input:}]}
\newcommand{\Output}{\item[\textbf{Output:}]}
\usepackage[noend]{algpseudocode}

\usepackage{cuted}
\usepackage{comment}
\usepackage{mathtools}
\DeclarePairedDelimiter\ceil{\lceil}{\rceil}

\usepackage{accents}
\newlength{\dhatheight}
\newcommand{\hhat}[1]{%
    \settoheight{\dhatheight}{\ensuremath{\hat{#1}}}%
    \addtolength{\dhatheight}{-0.35ex}%
    \hat{\vphantom{\rule{1pt}{\dhatheight}}%
    \smash{\hat{#1}}}}

\usepackage[bottom=0.94in, right=0.63in, left=0.63in, top=0.70in]{geometry}

\usepackage[compact]{titlesec}
\titlespacing{\section}{0pt}{*0.82}{*0.82}
\titlespacing{\subsection}{0pt}{*0.62}{*0.62}

\newcommand{\bd}[1]{\mathbf{#1}}

\algrenewcommand\algorithmicrequire{\textbf{Input:}}
\algrenewcommand\algorithmicensure{\textbf{Output:}}

\DeclareRobustCommand{\erase}{\bgroup\markoverwith{\textcolor{red}{\rule[.5ex]{2pt}{0.5pt}}}\ULon}

\def\BibTeX{{\rm B\kern-.05em{\sc i\kern-.025em b}\kern-.08em
    T\kern-.1667em\lower.7ex\hbox{E}\kern-.125emX}}

\begin{document}

\bstctlcite{IEEEexample:BSTcontrol}

\title{Low-complexity Frequency Domain Equalization for filtered-AFDM over General Physical Channels
\\
\thanks{This paper has been accepted by IEEE International Conference on Communications Workshops 2026.}
}

\author{\IEEEauthorblockN{Cheng~Shen, Chenyang Zhang and Jinhong~Yuan \big(\{cheng.shen1, chenyang.zhang1, j.yuan\}@unsw.edu.au\big)}
\IEEEauthorblockA{The University of New South Wales, Sydney, NSW, 2052, Australia\\
}
\vspace*{-8mm}
}

\maketitle

\begin{abstract}
Affine frequency division multiplexing (AFDM) has emerged as a promising waveform for high-mobility communications. However, its equalization remains a practical challenge under general physical channels with off-grid delay and Doppler effects. In this paper, we investigate frequency domain equalization for AFDM by considering a practical filtered-AFDM waveform. We analyze the input–output relations of filtered-AFDM across various domains and show that off-grid effects lead to severe inter-symbol interference in the DAFT domain, limiting the effectiveness of DAFT domain equalization. Motivated by the compactness of the frequency domain channel matrix in wideband systems, we propose a low-complexity two-stage frequency domain equalization scheme. Numerical results demonstrate that the proposed approach achieves performance close to full-block LMMSE equalization with significantly reduced computational complexity, and offers clear advantages over time domain equalization in wideband scenarios.
\end{abstract}

\section{Introduction}
The growing demand for reliable communications in high-mobility scenarios presents fundamental challenges to conventional orthogonal frequency division multiplexing (OFDM) systems, whose performance degrades severely in doubly selective channels with pronounced Doppler effects. This limitation has motivated the development of delay-Doppler domain multicarrier (DDMC) waveforms, represented by orthogonal time frequency space (OTFS) and orthogonal delay-Doppler division multiplexing ~\cite{Hadaniwcnc17,oddm}. By multiplexing information in the delay–Doppler (DD) domain, where the channel admits a more compact and stable description, DDMC enables improved robustness compared to OFDM over these channels.

More recently, affine frequency division multiplexing (AFDM) has been proposed as another waveform tailored to the same class of doubly selective channels~\cite{AFDM}. Unlike DDMC, AFDM multiplexes information symbols in the discrete affine Fourier transform (DAFT) domain using mutually orthogonal chirp signals. With appropriate selection of the post-chirp parameter, AFDM achieves error-rate performance comparable to OTFS while enabling channel estimation with lower overhead~\cite{AFDM}. Owing to these properties, AFDM has been further investigated in a variety of contexts and demonstrating great potential for emerging application scenarios.

Despite these advantages, equalization remains a key bottleneck for AFDM in practical deployments. Initial studies primarily focused on DAFT domain equalization, exploiting sparsity of the DAFT domain channel representation~\cite{AFDM,AFDM_LDL,AFDM_MP}. However, when the finite time duration of practical signals is taken into account, the resulting limited Doppler resolution gives rise to off-grid Doppler shifts. This effect undermines the assumed sparsity of the DAFT domain channel, leading to both performance degradation and increased computational burden for sparsity-based detectors. To address this issue, subsequent works have explored time domain (TD) equalization strategies, where off-grid Doppler spreading is inherently avoided and the equivalent TD channel matrix exhibits a (quasi-)banded structure determined by the maximum normalized delay~\cite{TZP_AFDM}.

To further advance AFDM toward practical applicability, it is necessary to consider additional physical constraints that arise in realistic systems. In particular, the finite signal bandwidth also induces off-grid delay effects, in conjunction with off-grid Doppler shifts. Such channels correspond to the so-called \emph{general physical channels} studied in DDMC literature~\cite{oddm_io}. Under these conditions, an explicit continuous-time formulation of AFDM becomes indispensable, yet remains largely unexplored in many existing AFDM studies. Moreover, insights from DDMC research indicate that TD equalization may become relatively inefficient in wideband systems, where large normalized delay spread lead to elevated computational complexity~\cite{adpt_2}. In these regimes, frequency domain (FD) equalization can offer a more favorable balance between performance and complexity~\cite{adaptive_FD,adpt_2}.

Motivated by these observations, this paper investigates frequency domain equalization for AFDM under general physical channels. Specifically,
\begin{itemize}
    \item We consider a practical filtered-AFDM (f-AFDM) waveform and derive its TD, FD, and DAFT domain input–output (IO) relations over general physical channels, highlighting the impact of off-grid delay and Doppler effects on each representation. We show that the DAFT domain equivalent channel exhibits widespread inter-symbol interference (ISI) due to the off-grid effect, similar to equivalent sampled DD domain channel in DDMC~\cite{oddm_io}. We further demonstrate through an example that time and frequency domains are more suitable to carry out equalization for their compactness, with the frequency domain being particularly advantageous for wideband systems.
    
   \item Based on the FD IO relation, we proposed a two-stage equalization scheme for f-AFDM. In particular, we employ block Cholesky factorization to obtain an initial estimate in the first stage. In the second stage, we perform cross domain equalization with a hard-decision fallback mechanism to further enhance error performance.
    \item Through numerical evaluations under practical system and channel parameters, we demonstrate that the proposed FD-based equalizer achieves performance close to full-block LMMSE equalization while substantially reducing computational complexity, and offers clear advantages over TD equalization in wideband settings.
\end{itemize}

\section{Preliminaries}

\subsection{Filtered-AFDM}
Consider the transmission of an AFDM frame \(\bd{x}\) within a nominal bandwidth \(B\) and a time frame \(T_f\), which contains \(N=BT_f\) symbols in DAFT domain. The discrete time domain representation of \(\bd{x}\) is obtained by performing an \(N\)-point inverse discrete affine Fourier transform (IDAFT)~\cite{AFDM},
\begin{equation}\label{eq_s}
    s[n] = \frac{1}{\sqrt{N}}\sum_{m=0}^{N-1}x[m]\phi^{H}(n,m),
\end{equation}
where \(0\leq n\leq N-1\), \(x[m]\) represents the \(m\)-th element of \(\bd{x}\), and \(\phi(n,m)\) denotes the DAFT kernel parameterized by the pre-chirp parameter \(c_2\) and the post-chirp parameter \(c_1\)~\cite{AFDM},
\begin{equation}\label{eq_DAFT_kernel}
    \phi(n,m) = e^{-j2\pi (c_{1}n^2+c_{2}m^2+\frac{nm}{N})}.
\end{equation}
Next, based on~\cite{AFDM}, an \(L_{\text{cpp}}\)-symbol-long chirp-periodic prefix (CPP) is added to \(\bd{s}\) to yield the discrete time domain transmitted signal, with elements in the CPP part specified by
\begin{equation} 
s_{\text{cpp}}[n] = s[N+n]e^{-j2\pi c_{1}(N^{2}+2Nn)},-L_{\text{cpp}}\leq n\leq-1.
\end{equation}
Here, we consider the cases \(2c_1N\in \mathbb{Z}\) and \(N\) is even. This essentially renders the CPP to coincide with the conventional CP, allowing for flexible interpretations that simplifies derivations for IO relations in the following section.  

In analogy to filtered-OFDM~\cite{fOFDM}, a prototype filter \(a(t)\) can be applied to \(\bd{s}_{\text{cpp}}\) to obtain a spectrally contained AFDM signal. 
This gives the filtered-AFDM (f-AFDM) with continuous time domain representation
\vspace{-1mm}
\begin{equation}\label{eq_tx_approx} 
    s(t)=\sum_{n=-L_{\text{cpp}}}^{N-1} {s}_{\text{cpp}}[n] a\left(t-nT_s\right).
\end{equation}

\subsection{Doubly Selective Channel}

We consider a doubly selective channel characterised by the Doppler-delay spreading function~\cite{bello} \vspace{-1mm}\begin{equation}\label{eq_h_dd}
h(\tau,\nu) = \sum _{p=1}^{P} h_{p} \delta (\tau -\tau _{p}) \delta (\nu -\nu_{p}),
\end{equation}
where \(P\) denotes the total number of \textit{physical} paths in the channel, and \(h_{p}\), \(\tau _{p}\), and \(\nu _{p}\) are the gain, delay, and Doppler shift of the \(p\)-th path, respectively. 
Given the system bandwidth 
and the signal time duration,
we obtain the normalized delay and Doppler shift associated with each path with respect to the DD resolution, \({\tau }_p=l_{p}T_{s}\) and \( {\nu }_p=k_{p}/T_{f}\), where \(l_p\) and \(k_p\) are \textit{real numbers}, and \(0\leq l_p\leq l_{\text{max}},\ |k_p|\leq k_{\text{max}}\), with \(l_{\text{max}}\) and \(k_{\text{max}}\) denoting the maximum normalized delay and Doppler shift of the channel, respectively. 



When \(s(t)\) in~\eqref{eq_tx_approx} propagates through a channel characterised by~\eqref{eq_h_dd}, the received continuous-time signal becomes~\cite{bello}
\vspace{-1mm}
\begin{equation} \label{Equ:rt}
\tilde{r}(t)=\sum_{p=1}^{P}h_p\!\!\sum_{n=-L_{\text{cpp}}}^{N-1}\!\!\!s[n]a(t{-}nT_s{-}\tau_p) e^{j2\pi\nu_p(t-\tau_p)}+w(t),
\end{equation}
where \(w(t)\sim\mathcal{CN}(0,\sigma^2)\)  represents the additive white Gaussian noise (AWGN).

\section{IO relations for f-AFDM in various domains}\label{sec:3}
At receiver, \(r(t)\) is passed through an \(a(t)\)-based matched filter, which gives the filtered signal~\cite{oddm_io}
\begin{equation*}
    {r}(t)=\sum_{p=1}^{P}\tilde{h}_p\!\!\sum_{n=-L_{\text{cpp}}}^{N-1}s[n]g(t-nT_s-\tau_p) e^{j2\pi\nu_p(t-\tau_p)}+{w}(t),
\end{equation*}
where \(\tilde{h}_{p}=h_p e^{-j2\pi \nu_p\tau_p}\), and \(g(t)=a(t)\ast a^{*}(-t)\) is the composition filter, with \((\cdot)^{*}\) denoting complex conjugation and \(\ast\) denoting convolution. 
\subsection{Time domain IO relation}
Sampling at \(t=nT_s\) for \(0\leq n\leq N-1\) 
gives the discrete time domain representation of the received signal 
\begin{align}\label{eq_r_samp}
    \!\!&r[n]=\sum_{p=1}^{P}\tilde{h}_p\!\!\!\sum_{n'=-L_{\text{cpp}}}^{N-1}\!\!\!\!s_{\text{cpp}}[n']g((n{-}n')T_s-\tau_p) e^{j2\pi\nu_pnT_s}+w[n]\nonumber\\
    &=\sum_{p=1}^{P}\tilde{h}_p\sum_{n'=0}^{N-1}\!s[n']g((n{-}n')_N T_s-\tau_p) e^{j2\pi\frac{k_p n}{N}}+w[n],
\end{align}
where we interpret the prefix as CP so that \(s_{\text{cpp}}[n-d]=s[(n-d)_N]\), with \((\cdot)_N\) denoting modulo \(N\). In matrix form, we have 
\vspace{-1mm}
\begin{equation}\label{Eq_td_IO_matrix}
    \bd{r}=\bd{H}\bd{s}+\bd{w},
\end{equation}
where \(\bd{H}\) denotes the \(N\times N\) TD channel matrix, and \(\bd{s}\), \(\bd{r}\), and \(\bd{w}\) denote the \(N\times 1\) TD transmitted, received, and noise signal vectors, respectively. Based on~\eqref{eq_r_samp}, we can obtain
\vspace{-1mm}
\begin{align}
    H[n,n']=\sum_{p=1}^{P}\tilde{h}_p g((n{-}n')_N T_s-\tau_p) e^{j2\pi\frac{k_p n}{N}},
\end{align}
Furthermore, without loss of generality, we suppose that \(g(t)\) has an effective support \([0,DT_s]\) with \(D\) denoting a positive integer. This suggests that contribution of the \(p\)-th path to \(\bd{H}[n,n']\) is non-trivial when \(\lceil l_p\rceil\leq (n-n')_D\leq D+\lfloor l_p\rfloor\). Given the range of \(l_p\), \(\bd{H}\) is approximately cyclically banded with a lower bandwidth \(D+\ceil{l_{\text{max}}}\).

\subsection{Frequency domain IO relation}\label{subsec:FDIO}
We first substitute \(d=n-n'\) into~\eqref{eq_r_samp} while recognizing the finite support of \(g(t)\), which gives
\begin{equation}\label{eq_r_dis}
    r[n]=\sum_{p=1}^{P}\tilde{h}_{p}\!\!\sum_{d=\lceil l_p\rceil}^{D+\lfloor l_p\rfloor}\!\!s[(n-d)_N]g(dT_s-\tau_p)e^{j2\pi\frac{k_p n}{N}} +w[n].
\end{equation}
Let \(\dot{\bd{s}}\) and \(\dot{\bd{r}}\) denote the discrete frequency domain representations of the transmitted and received signal, respectively. Performing \(N\)-point DFT on both sides of~\eqref{eq_r_dis} and writing \(\bd{s}\) as the IDFT of \(\dot{\bd{s}}\), we obtain 
\begin{align}\label{eq_f_IO}
    \dot{r}[\dot{n}]=\sum_{p=1}^{P}\tilde{h}_p\sum_{\Bar{n}=0}^{N-1}\dot{g}_p[\Bar{n}]\dot{s}[\Bar{n}]\Omega(\Bar{n}-\dot{n}+k_p)+\dot{w}[\dot{n}],
\end{align}
where \(\dot{g}_p[\Bar{n}]=\sum_{d=\lceil l_p\rceil}^{D+\lfloor l_p\rfloor}g(dT_s-\tau_p) e^{-j\frac{2\pi}{N}\Bar{n}d}\), \(\dot{w}[\dot{q}]\) is the noise term in FD, and
\vspace{-2mm}
\begin{equation}
    \Omega(k)=\sum_{n=0}^{N-1}e^{j\frac{2\pi}{N}nk}
\end{equation}
denotes the Dirichlet function. We express~\eqref{eq_f_IO} in matrix form
\begin{equation} \label{eq_fd_IO_matrix}
    \dot{\bd{r}}=\bd{\dot{H}}\dot{\bd{s}}+\dot{\bd{w}},
\end{equation}
where \(\bd{\dot{H}}\) denotes the \(N\times N\) FD channel matrix, with
\begin{equation}\label{eq_H_f}
    \dot{\bd{H}}[\dot{n},\Bar{n}]=\sum_{p=1}^{P}\tilde{h}_p\dot{g}_p[\Bar{n}]\Omega(\Bar{n}-\dot{n}+k_p). 
\end{equation}
Moreover, we note that \(|\Omega(\Bar{n}+k_p-\dot{n})|\) can be relatively small when \(|\Bar{n}+k_p-\dot{n}|>\gamma\), where \(\gamma\) is a positive integer approximating half of the size of the Dirichlet function's support. Then, considering the range of \(k_p\), \(\dot{\bd{H}}\) is approximately cyclically banded with an upper and lower bandwidth \(\gamma+\lceil k_{\text{max}}\rceil\).

\subsection{DAFT domain IO relation}
To obtain a concise IO relation, we now interpret the prefix as CPP, so that in~\eqref{eq_r_dis},
\begin{equation}\label{eq_s_cpp_d}
    s[(n-d)_N]=\sum_{m=0}^{N-1}x[m]e^{-j2\pi (c_{1}(n-d)^2+c_{2}m^2+\frac{(n-d)m}{N})}
\end{equation}
holds for all \(0\leq d\leq L_{\text{cpp}}\) and \(0\leq n\leq N-1\). Then, we perform \(N\)-point DAFT on both sides of~\eqref{eq_r_dis} and substituting~\eqref{eq_s_cpp_d}. With some simple rearrangement, we obtain  
\begin{align}\label{eq_DAFT_IO}
    &y[\dot{m}]=e^{-j2\pi c_2\dot{m}^2}\sum_{p=1}^{P}\tilde{h}_p\sum_{d=\lceil l_p\rceil}^{D+\lfloor l_p\rfloor}\!\!g(dT_s-\tau_p)e^{j2\pi c_1 d^2}\nonumber\\
    &\times\sum_{m=0}^{N-1}x[m]e^{j2\pi(c_2m^2-\frac{dm}{N})}\Omega(m-\dot{m}-2c_1 Nd+k_p)+\ddot{w}[m],
\end{align}
for \(0\leq \dot{m}\leq N-1\). Equivalently, with \(\bd{\ddot{H}}\in\mathbb{C}^{N\times N}\), \(\bd{y},\ddot{\bd{w}}\in \mathbb{C}^{N\times 1}\) denoting the DAFT domain channel matrix, received signal and noise vectors, respectively, we have
\begin{equation}
    \bd{y}=\bd{\ddot{H}}\bd{x}+\ddot{\bd{w}},
\end{equation}
\begin{align}\label{eq_H_DAFT}
    &\ddot{H}[\dot{m},m]=e^{-j2\pi c_2\dot{m}^2}\sum_{p=1}^{P}\tilde{h}_p\sum_{d=\lceil l_p\rceil}^{D+\lfloor l_p\rfloor}\!\!g(dT_s-\tau_p)e^{j2\pi c_1 d^2}\nonumber\\
    &\times e^{j2\pi(c_2m^2-\frac{dm}{N})}\Omega(m-\dot{m}-2c_1 Nd+k_p).
\end{align}

\underline{Discussions:} From~\eqref{eq_H_DAFT}, a DAFT domain transmitted symbol \(x[m]\) due to the \(p\)-th path can be viewed as spreading into approximately \(D\) clusters in the DAFT domain received signal, with each cluster centres around \(y[m{+}2c_1 Nd]\) and weighted by \(\tilde{h}_p g(dT_s-\tau_i)\) for \(\lceil l_p\rceil \leq d\leq D+\lfloor l_p\rfloor\). Within each cluster, the symbol further spread out following the Dirichlet function centreing around \(y[m{+}2c_1 Nd{+}\lfloor k_p\rceil]\). Thus, the footprint of the \(p\)-th path has its main lobe (highest magnitude) on the \((2c_1N\lfloor l_p\rceil-\lfloor k_p\rceil)\)-th cyclic diagonal of \(\ddot{\bd{H}}\) with a weight \(g(dT_s-\tau_i)\Omega(k_p-\lfloor k_p\rceil)\). Also, given the range of \(l_p\) and \(k_p\), overall, \(x[m]\) can spread over \(\lambda=2c_1 N(D+\ceil{l_{\text{max}}})+2(\gamma+\lceil k_{\text{max}}\rceil)\) DAFT domain symbols, or equivalently, \(\ddot{\bd{H}}\) has an approximate bandwidth \(\lambda\). Note that an important criterion for the \(c_1\) selection that maximizes diversity is to select \(c_1\) to separate the footprint of each path in \(\ddot{\bd{H}}\)~\cite{AFDM}. However, with a widespread footprint due to off-grid effect, this will cause \(\ddot{\bd{H}}\) to have a bandwidth \(\lambda\) significantly higher than that for \(\bd{H}\) and \(\dot{\bd{H}}\). On the other hand, the minimal DAFT domain spread is attained when \(c_1=0\), so that all clusters overlap and \(\lambda=2(\gamma+\lceil k_{\text{max}}\rceil)+1\), in which case the DAFT domain degenerate to frequency domain. In this sense, frequency domain can be viewed as the minimal spreading DAFT domain.

\subsection{A demonstrative example}\label{subsec:example}
A demonstrative example is provided by considering a system with approximately \(B=7.68\) MHz and time frame duration \(T_f=66.7\mu s\), containing \(N=512\) symbols. Root raised cosine (RRC) with roll-off factor 0.1 is employed as the prototype filter \(a(t)\). For AFDM, we set \(c_1=1/N\). We consider a channel with delays of the paths specified by the EVA model, carrier frequency \(f_c=6\) GHz, maximum user equipment (UE) speed \(v_{\mathrm{max}}=500\)km/h, and the Doppler shift generated using Jakes' model. These settings characterize a typical wideband system operating in high mobility environment. 

\begin{figure*}[t]
\centering
    \includegraphics[width=0.98\textwidth]{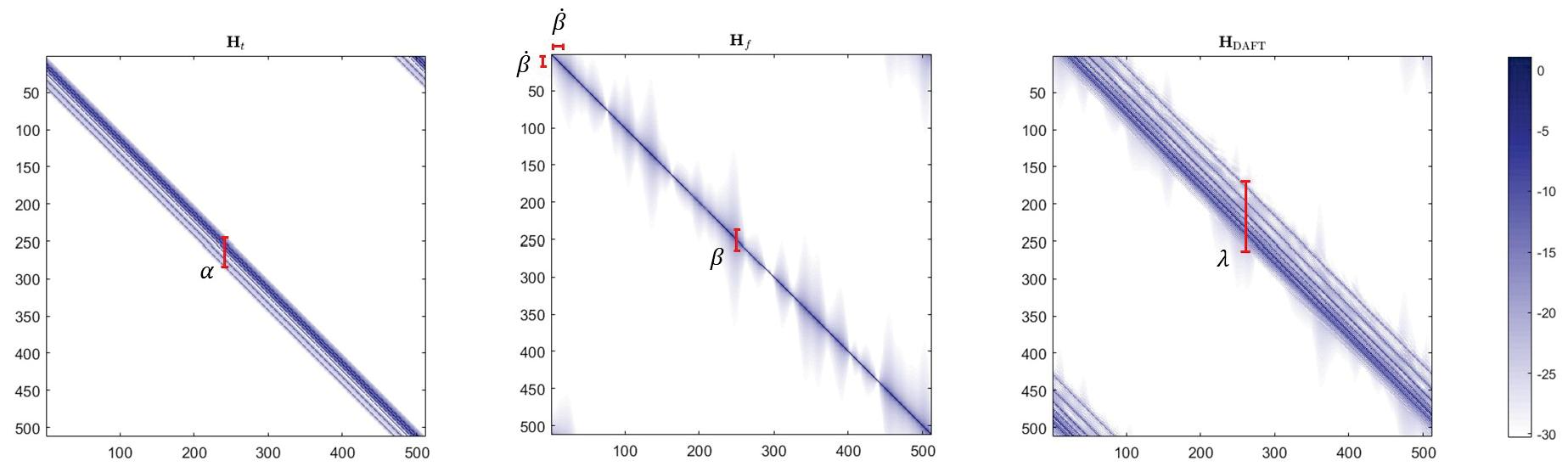}
\vspace{-4mm}
\caption{An example for channel matrices in time (left), frequency (middle), and DAFT (right) domains. The system has \(B=7.68\text{MHz}\), \(T_f=66.7\mu s\), corresponding to \(N=512\), and \(f_c=6\text{GHz}\). For the realization in display, the channel has 9 taps with power \((-3.6,-5.9,-10.5,-4.2,-5.0,-6.8,-6.5,-9.2,-11.7)\) dB, delays specified by EVA model, and associated moving speed \((153,-472,472,-380,3,189,496,482,-486)\) km/h. Matrix entries with magnitude below -30dB display as white. \(\alpha\) and \(\beta\) are the cyclic bandwidth used in band approximation of \(\bd{H}\) and \(\dot{\bd{H}}\) during equalization, respectively, with \(\beta=2\dot{\beta}+1\).}
\label{fig:ch_mtx}
\vspace{-4mm}
\end{figure*}

Fig.~\ref{fig:ch_mtx} shows the entries in \(\bd{H}(\bd{H}_t)\), \(\dot{\bd{H}}(\bd{H}_f)\) and \(\ddot{\bd{H}}(\bd{H}_{\text{DAFT}})\) with magnitude greater than -30dB. As shown in the right subfigure, the DAFT domain channel matrix \(\ddot{\bd{H}}\) exhibits a high bandwidth, aligning with the analytical insights in previous discussions. In particular, the off-grid delay and Doppler shift lead to widespread ISI following the composition filter \(g(t)\) and Dirichlet function \(\Omega(k)\), which hinders the DAFT domain sparsity. In comparison, TD and FD channel matrices \(\bd{H}\) and \(\dot{\bd{H}}\) are more compact. However, due to a higher normalized delay spread than Doppler spread in the wideband system, \(\bd{H}\) in this example has a bandwidth around 20, whereas the significant entries of \(\dot{\bd{H}}\) mostly centres around the main diagonal. More generally, \(l_{\text{max}}\) is typically on the order of several tens for a system of similar bandwidth, whereas for sub-6GHz signals with \(T_f<1\mathrm{ms}\), \(k_{\text{max}}\) would not exceed \(3\) and is thus far below \(l_{\text{max}}\). These observations reflect that FD equalization can potentially enjoy lower complexity thanks to a more compact channel matrix comparing to time and DAFT domains.
\section{Frequency domain Equalization for f-AFDM}

Leveraging the compact structure of the FD channel matrix \(\dot{\bd{H}}\), we propose a FD equalization scheme for f-AFDM following a two-stage approach. The first stage provides a low-complexity initial equalization, and the second stage performs iterative detection to enhance error performances.

\subsection{Stage 1: Cholesky factorization based LMMSE}
Given the quasi-banded structure of FD channel matrix \(\dot{\bd{H}}\) as per Section~\ref{subsec:FDIO}, we first obtain its band approximation
\begin{equation}
    \Breve{\bd{H}} = \dot{\bd{H}} \odot \boldsymbol{\Xi} ,
\end{equation}
where \(\odot\) denotes Hadamard multiplication, \(\boldsymbol{\Xi} =\sum_{q=-\dot{\beta}}^{\dot{\beta}}\bd{\Pi}^{q}\), \(\boldsymbol{\Pi}=\mathrm{circ}([0,0,\dots,0,1])\in \mathbb{C}^{N\times N}\), with \(\mathrm{circ}(\bd{v})\) denoting the circulant matrix whose first row is \(\bd{v}\). Then, the approximate FD IO relation is given by
\begin{equation}\label{eq_freq_IO_approx}
    \dot{\bd{r}}=\Breve{\bd{H}}\dot{\bd{s}}+\dot{\bd{w}}.
\end{equation} 
where \(\Breve{\bd{H}}\) is cyclically banded with an upper and lower bandwidth \(\dot{\beta}\), or a total bandwidth \(\beta=2\dot{\beta}+1\). Note that \(\dot{\beta}\) is subject to design choice, with a larger \(\dot{\beta}\) leading to more accurate equalization yet higher computational complexity. 
Based on this, an initial LMMSE estimate of \(\dot{\bd{s}}\) can be obtained as
\begin{equation}\label{eq_LMMSE}
    \hat{\dot{\bd{s}}} = \Breve{\bd{H}}^{H}(\Breve{\bd{H}}\Breve{\bd{H}}^H+\sigma^2\bd{I})^{-1}\dot{\bd{r}}.
\end{equation}
Direct computation of~\eqref{eq_LMMSE} requires \(\mathcal{O}(N^3)\) complex multiplications (CMs) (for simplicity, we also count complex divisions as CMs). Here, we leverage the band structure of \(\Breve{\bd{H}}\) and Cholesky factorization to reduce the complexity to \(\mathcal{O}(N\beta^2)\). Specifically, since \(\Breve{\bd{H}}\) is cyclically banded, in~\eqref{eq_LMMSE}, the matrix to be inverted
\vspace{-1mm}
\begin{equation}\label{eq_G_mtx}
    \bd{G}=\Breve{\bd{H}}\Breve{\bd{H}}^H+\sigma^2\bd{I}
\end{equation}
is cyclically banded with upper and lower bandwidth \(\beta\), with~\eqref{eq_G_mtx} itself taking \(\frac{1}{2}(\beta^2{+}3\beta{+}1)N\) CMs~\cite{matrix_comp}. Note that this differs from TD processing proposed in~\cite{TZP_AFDM}, where zero-padding can be used to obtain a strictly banded \(\bd{G}\) without additional spectral efficiency loss. Here, special treatment is requires to handle the non-trivial terms on the top-right and bottom-left corners of \(\bd{G}\).

We note that \(\bd{G}\) is Hermitian positive definite (HPD). Then, its Cholesky factorization gives \(\bd{G}=\bd{L}\bd{L}^H\) with \(\bd{L}\) being lower-triangular. Partitioning each matrix yields~\cite{matrix_comp}
\begin{align}\label{eq_LU_partition} 
\underbrace {\begin{bmatrix} \bd {G}^{(1)}_{Q\times Q} & {\bd{G}}^{(2)}_{ Q\times \beta } \\ {\bd{G}}^{(3)}_{ \beta \times Q} & {\bd{G}}^{(4)}_{\beta \times \beta } \end{bmatrix}}_{ \bd{G}}
= 
\underbrace {\begin{bmatrix} \bd {A}_{Q\times Q} & \bd {0}_{ Q\times \beta } \\ {\bd{B}}_{ \beta \times Q} & {\bd{C}}_{ \beta \times \beta } \end{bmatrix}}_{ \bd {L}}
\times 
\underbrace {\begin{bmatrix} \bd {A}^{H}_{Q\times Q} & \bd {\bd{B}}^{H}_{\mathrm{ Q\times \beta }} \\ \bd {0}_{\mathrm{ \beta \times Q}} & \bd {C}^{H}_{\mathrm{ \beta \times \beta }} \end{bmatrix}}_{\bd{L}^H}
\end{align}
where \(\bd{A}\) and \(\bd{C}\) are lower triangular. Thus, \(\bd{L}\) can be found by solving for \(\bd{A}\), \(\bd{B}\), and \(\bd{C}\) in~\eqref{eq_LU_partition}. We first use the relation
\vspace{-1mm}
\begin{equation}
    \bd {A}\bd{A}^{H}=\bd{G}^{(1)}.
\end{equation}
to find \(\bd{A}\). Note that \(\bd{G}^{(1)}\) is HPD since it is the leading principle block of \(\bd{G}\). Also, it is banded with a upper and lower bandwidth \(\beta\). Thus, \(\bd{A}\) can be obtained with a band Cholesky factorization on \(\bd{G}^{(1)}\), which is a lower-banded matrix with bandwidth \(\beta\). This process takes \(N(\frac{1}{2}\beta^2+\beta)\) CMs. Next, \(\bd{B}\) can be found by solving
\vspace{-1mm}
\begin{equation}
    \bd{B}^{H} = \bd{A}^{-1}\bd{G}^{(2)}
\end{equation}
with a band-forward substitution, which requires about \((N-\beta)\beta^2\) CMs.
Furthermore, we solve \(\bd{C}\) from
\begin{equation}
    \bd {C}\bd{C}^{H}=\bd{G}^{(4)}-\bd {B}\bd{B}^{H}.
\end{equation}
Since \(\bd{G}^{(4)}{-}\bd {B}\bd{B}^{H}=\bd{G}^{(4)}-\bd{G}^{(3)}\left(\bd{G}^{(1)}\right)^{-1}\bd{G}^{(2)}\) is the Schur complement of \(\bd{G}\) with respect to \(\bd{G}^{(1)}\), it is also HPD. Hence, \(\bd{C}\) corresponds to the Cholesky factorization of \(\bd{G}^{(4)}-\bd {B}\bd{B}^{H}\), requiring approximately \(\frac{1}{2}\beta^3+\beta^2-\frac{1}{3}\beta\) CMs, including the computation of \(\bd {B}\bd{B}^{H}\). After that, we have obtained \(\hat{\dot{\bd{s}}}^{(0)} = \Breve{\bd{H}}^H(\bd{L}^{H})^{-1}\bd{L}^{-1}\dot{\bd{r}}\),
which can be solved with one forward substitution, one backward substitution, and a matrix multiplication, taking \(5N\beta-4\beta^2-2\beta\) CMs in total.

In addition, we also evaluate the error variance associated with the estimate \(\hat{\dot{s}}[n]\) for each \(0\leq n\leq N{-}1\), which is given by the diagonal elements of the covariance matrix from the LMMSE estimate~\cite{otfs_xdomain},
\begin{equation}\label{eq_freq_var}
    e^{(0)}[n]=1-||\bd{L}^{-1}\Breve{\bd{h}}_n||^2,
\end{equation}
where \(\Breve{\bd{h}}_n=\Breve{\bd{H}}[n{-}\dot{\beta}:n{+}\dot{\beta},n]\). Furthermore, to obtain a sufficiently accurate variance, it suffices to evaluate only the \([n{-}\dot{\beta},n{+}\dot{\beta}]\)-th elements of \(\bd{L}^{-1}\Breve{\bd{h}}_n\) during the forward substitution~\cite{TZP_AFDM}. Hence, calculating~\eqref{eq_freq_var} for all \(0\leq n\leq N{-}1\) takes \(N(\beta+1)^2\) CMs.

In summary, the total number of CMs required to obtain the initial LMMSE estimate and the associated error variance is
\begin{equation}
    \eta_0=N(3\beta^2+11\beta+\frac{5}{2})-\frac{1}{2}\beta^3-3\beta^2-\frac{2}{3}\beta.
\end{equation}

\subsection{Stage 2: Cross-domain iterative MMSE detection}


We propose a cross-frequency-and-DAFT-domain iterative MMSE detection algorithm for stage 2. Apart from FD processing, we further introduce a hard-decision fallback mechanism to accelerate convergence and reduce computational complexity, comparing to its TD counterpart~\cite{TZP_AFDM}.

In the \(i\)-th iteration with \(1\leq i\leq i_{\text{max}}\), we first obtain the intermediate estimate on the DAFT domain transmitted symbols \(\hhat{\bd{x}}\) and its associated variance vector \(\boldsymbol{\varepsilon}\) based on \(\hat{\dot{\bd{s}}}^{(i-1)}\) and \(\bd{e}^{(i-1)}\),
\begin{align}\label{eq_var_soft}
    \!\!\!\hhat{\bd{x}} = \bd{\Phi}\bd{F}^H\hat{\dot{\bd{s}}}^{(i-1)}, \
    {\boldsymbol{\varepsilon}} = \bd{1}_N\cdot \varepsilon, 
\end{align}
where \(\bd{\Phi}\in \mathbb{C}^{N\times N}\) is the DAFT matrix, and the symbol-wise variance \(\varepsilon[m]=\varepsilon=\frac{1}{N}\sum_{n=0}^{N-1}{e}^{(i-1)}[n]\) are approximated to be identical for all DAFT domain symbols \(\hhat{x}[m]\) (\(0\leq m\leq N-1\))~\cite{otfs_xdomain}. Based on this, for each possible constellation alphabet \(a\in \mathcal{X}\) that \(x[m]\) may represent, the normalized conditional probability for the intermediate estimate \(\hhat{x}[m]\) is given by
\begin{equation}\label{eq_norm_prob}
    {\Bar{\text {Pr}}}\{\hhat{x}[m]|x[m]=a\} = \frac{\text {exp}\left({-{|\hhat{x}[m]{-}a|^{2}}/{\varepsilon[m]}}\right)}{\sum_{\dot{a}\in\mathcal{X}}\text {exp}\left({-{|\hhat x[m]{-}\dot{a}|^{2}}/{\varepsilon[m]}}\right)}.
\end{equation}
The soft posterior mean of the DAFT domain symbols \(\tilde{\bd{x}}^{(i-1)}\) and the associated soft posterior variance \(\tilde{\bd{v}}^{(i-1)}\) can thus be evaluated, respectively, as
\begin{align} 
&{\tilde {x}}^{(i-1)}[m]= \sum _{a\in \mathcal{X}}{\Bar{\text{Pr}}}\{\hhat {x}[m]|x[m]=a\}\times a, \label{eq_DAFT_soft_est} \\
&\tilde{v}^{(i-1)}[m]= \sum _{a\in \mathcal{X}}{\Bar{\text{Pr}}}\{\hhat{x}[m]|x[m]{=}a\}\times |a{-}{\tilde {x}}^{(i-1)}[m]|^{2}.\label{eq_DAFT_soft_var} 
\end{align}
After that, the posterior statistics \(\tilde{\bd{x}}^{(i-1)}\) and \(\tilde{\bd{v}}^{(i-1)}\) is transformed into frequency domain to serve as the prior mean and variance for \(\dot{\bd{s}}\) in the current iteration,
\begin{align}\label{eq_freq_mean}
    \!\!\!\Bar{\dot{\bd{s}}}^{(i)} = \bd{F}\bd{\Phi}^H\tilde{\bd{x}}^{(i-1)}, \
    {\boldsymbol{\psi}}^{(i)} = \bd{1}_N\cdot \mu^{(i-1)},
\end{align}
where \(\mu^{(i-1)}\) is the mean of elements in \(\bd{v}^{(i-1)}\). The iteration stops here if $|\tilde{\bd{x}}^{(i)}{-}\tilde{\bd{x}}^{(i-1)}|<\varsigma$, where \(\varsigma\) is the halt threshold.

We next perform symbol-by-symbol estimation using the updated statistics. To estimate \(\dot{s}[n]\), we use the local IO relation
\begin{align}\label{eq_int_cancel}
     \dot{\bd{r}}_{n}=\bd{\Breve{H}}_{n}\bd{\dot{s}}_{n}+\bd{\dot{w}}_{n},
\end{align}
where \(\bd{\Breve{H}}_{n}{=}\bd{\dot{H}}[n{-}\dot{\beta}:n{+}\dot{\beta},n{-}2\dot{\beta}:n{+}2\dot{\beta}]\), \(\bd{\dot{r}}_n{=}\dot{r}[n{-}\dot{\beta}:n{+}\dot{\beta}]\), \(\bd{\dot{s}}_{n}=\bd{\dot{s}}[n{-}2\dot{\beta}:n{+}2\dot{\beta}]\) and \(\bd{\dot{w}}_{n}=\dot{\bd{w}}[n{-}2\dot{\beta}:n{+}2\dot{\beta}]\). Note that all the indexes are taken modulo-\(N\). 
Based on this, the local LMMSE estimate on \(\dot{s}[n]\) is given by
\begin{align}\label{eq_local_MMSE}
    \!\!\!&\hat{\dot{s}}^{(i)}[n] = \nonumber\\
    &\Bar{\dot{s}}^{(i)}[n]{+}\psi^{(i)}[n]\Breve{\bd{h}}_{n}^H\left(\Breve{\bd{H}}_{n}{\boldsymbol{\Psi}}^{(i)}_{n}\Breve{\bd{H}}_{n}^H{+}\sigma^2\bd{I}\right)^{-1}\!\!(\bd{\dot{r}}_{n}{-}\bd{\Breve{H}}_{n}\Bar{\dot{\bd{s}}}^{(i)}_{n}),
\end{align}
where \(\Breve{\bd{h}}_{n}\) represents the \(\beta\)-th column of \(\Breve{\bd{H}}_{n}\) and \({\boldsymbol{\Psi}}_{n}{:=}\text{diag}({\boldsymbol{\psi}}_{n})\) is the local covariance matrix. Here, we use only the extrinsic information by temporarily assuming \(\Bar{\dot{s}}^{(i)}[n]=0\) and \(\psi^{(i)}[n]=1\), so that
\begin{align}
 &\Bar{\dot{\bd{s}}}^{(i)}_{n}{:=}
 \left[\Bar{\dot{s}}^{(i)}[n{-}2\dot{\beta}:n{-}1]^T\!,\ 0,\ \Bar{\dot{s}}^{(i)}[n{+}1:n{+}2\dot{\beta}]^T\right]^T \label{eq_mean_local_vec}\\
 &{\boldsymbol{\psi}}^{(i)}_{n}{:=}\left[\psi^{(i)}[n{-}2\dot{\beta}:n{-}1]^T\!,\ \!1,\!\ \psi^{(i)}[n{+}1:n{+}2\dot{\beta}]^T\right]^T \label{eq_var_local_vec}
\end{align}
Also, the variance of estimation error for \(\hat{\dot{s}}^{(i)}[n]\) is
\begin{equation}\label{eq_local_err}
    e^{(i)}[n] = 1-\Breve{\bd{h}}_{n}^H\left(\Breve{\bd{H}}_{n}{\boldsymbol{\Psi}}_{n}\Breve{\bd{H}}_{n}^H+\sigma^2\bd{I}\right)^{-1}\Breve{\bd{h}}_{n}.
\end{equation}
The estimates in~\eqref{eq_local_MMSE} and error variance in~\eqref{eq_local_err} are stacked to form \(\hat{\dot{\bd{s}}}^{(i)}\) and \(\bd{e}^{(i)}\), respectively, which will be used in the next iteration.
\begin{algorithm}
  \caption{Proposed Two-stage FD Equalization}\label{alg:t_soft}
  \begin{algorithmic}[1]
    \small
    \Input $\dot{\bd{r}}$, $\dot{\bd{H}}$, $\sigma^2$
    \State Calculate initial FD estimate $\hat{\dot{\bd{s}}}^{(0)}$ and error $\bd{e}^{(0)}$ using~\eqref{eq_G_mtx}-\eqref{eq_freq_var}.
    \For{$i=1$ to $i_{\text{max}}$}
      \For{$n=0$ to $N-1$}  
          \State Obtain $\hhat{\bd{x}}$ and \(\varepsilon\) from~\eqref{eq_var_soft}.
          \State \parbox[t]{\dimexpr\linewidth-\algorithmicindent-2em\relax}{%
            Evaluate FD prior mean and variance using~\eqref{eq_norm_prob}-\eqref{eq_freq_mean}.
            If $\varepsilon<0.1$, replace~\eqref{eq_norm_prob} and~\eqref{eq_DAFT_soft_est}
            with \(\tilde{{x}}^{(i-1)}[n]=\arg \min_{a\in \mathcal{X}} |a-\hhat{x}[n]|\);
            \(\bd{\psi}^{(i)}=\bd{0}_N\) in~\eqref{eq_freq_mean}.}
          \State Exit if $|\tilde{\bd{x}}^{(i)}-\tilde{\bd{x}}^{(i-1)}|<\varsigma$.
          \State \parbox[t]{\dimexpr\linewidth-\algorithmicindent-2em\relax}{Obtain FD estimate $\hat{\dot{\bd{s}}}^{(i)}$ and error $\bd{e}^{(i)}$ from~\eqref{eq_local_MMSE}-\eqref{eq_local_err}. If $\varepsilon<0.1$, use~\eqref{eq_hard_simp} in~\eqref{eq_local_MMSE} and~\eqref{eq_local_err}.}
        \EndFor
    \EndFor
\Output $\hat{{\bd{x}}}=\arg \min_{\bd{a}\in \mathcal{X}^N} |\bd{a}-\tilde{\bd{x}}|$
  \end{algorithmic}
\end{algorithm}

\underline{Hard-decision fallback:} Directly implementing Eqs.~\eqref{eq_var_soft}-\eqref{eq_local_err} requires
\(\eta_{\text{soft}}=2N\log_{2}{N}+\frac{N}{12}(2\beta^3+45\beta^2+109\beta)+11N+\frac{17N}{4}|\mathcal{X}|\)
CMs per iteration~\cite{TZP_AFDM}, with the dominating \(\beta^3\) and \(\beta^2\) terms coming from the matrix inversion and multiplications in~\eqref{eq_local_MMSE} and~\eqref{eq_local_err}. However, when the variance of intermediate estimate \(\varepsilon\) from the previous estimate is small, the softmax-based evaluations in~\eqref{eq_norm_prob} and~\eqref{eq_DAFT_soft_est} reduces to the hard-decision \(\tilde{{x}}^{(i-1)}[n]=\arg \min _{a\in \mathcal{X}} |a-\hhat{x}[n]|\) for \(0{\leq} n{\leq} N{-}1\). Also,  this implies full confidence with the estimate and therefore assumes \(\tilde{{v}}^{(i-1)}[n]=0\) in~\eqref{eq_DAFT_soft_est}, leading to \({\boldsymbol{\psi}}^{(i)}=\bd{0}_N\) in~\eqref{eq_freq_mean}. Using this in~\eqref{eq_var_local_vec} and subsequently in~\eqref{eq_local_MMSE}, we have
\begin{align}\label{eq_hard_simp}
\!\!\Breve{\bd{h}}_{n}^H\!\left(\Breve{\bd{H}}_n{\boldsymbol{\Psi}}^{(i)}_n\Breve{\bd{H}}_n^H{+}\sigma^2\bd{I}\right)^{-1}\!\!\!=\left(\Breve{\bd{h}}_{n}^H\Breve{\bd{h}}_{n}{+}\sigma^2\bd{I}\right)^{-1}\!\Breve{\bd{h}}_{n}^H =\!\frac{\Breve{\bd{h}}_{n}^H}{{|\Breve{\bd{h}}_{n}|^2{+}\sigma^2}}.
\end{align}
Thus, without matrix inversion, the complexity per iteration reduces to \(\eta_{\text{hard}}=2N\log_2{N}+N(2\beta+7+|\mathcal{X}|/2)\) CMs. We spare step-by-step counting due to space limitations.

The process of the two-stage FD equalization is summarised in Alg.~\ref{alg:t_soft}. The total number of CMs required is \(\eta_0+i_{\text{hard}}\eta_{\text{hard}}+i_{\text{soft}}\eta_{\text{soft}}\), where \(i_{\text{hard}}\) and \(i_{\text{soft}}\) are the number of iterations performed using hard and soft decisions, respectively.

\section{Numerical Results}

In this section, we numerically evaluate the error performance and computational complexity of the proposed FD two-stage equalization with hard-decision fallback to other existing and related schemes. The system and channel settings follow from Section~\ref{subsec:example}, with at least \(10^8\) random realizations of channels tested or 100 bit errors collected at each \(E_s/N_0\) value.

\begin{figure}
\centering
    \includegraphics[width=3.0in]{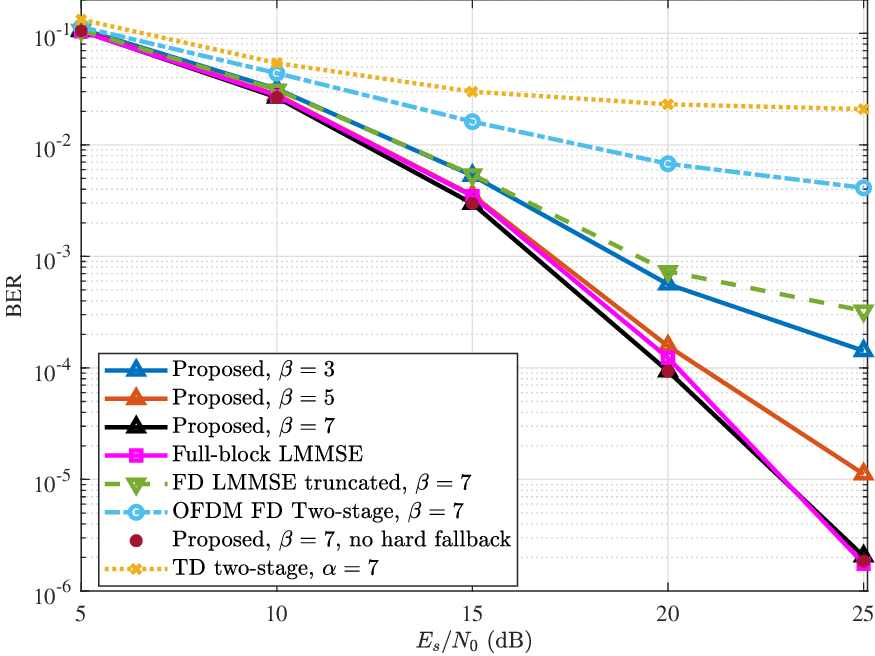}
    \vspace{-2mm}
\caption{Error performance for different equalization schemes for f-AFDM.}
\label{fig:eva}
\end{figure}

The comparative performance of the proposed scheme and other schemes are shown in Fig.~\ref{fig:eva}. First, the proposed scheme is evaluated under different values for the approximated FD channel matrix cyclic bandwidth \(\beta\), where a larger \(\beta\) indicates a less aggressive band approximation in pursuit of better error performance. As can be seen, increasing \(\beta\) from 3 to 7 leads to a quick drop of BER at high \(E_s/N_0\). Meanwhile, the proposed scheme with \(\beta=7\) also approaches the performance of full-block LMMSE at a lower computational cost, whereas the latter requires \(\mathcal{O}(N^3)\) CMs.

We then focus on schemes where band approximation on the channel matrix are performed with \(\beta=7\), indicated with different line styles. Compared to LMMSE based on band approximation \(\beta=7\) (corresponding to perform stage 1 of our equalizer only, shown as the green trace in Fig.~\ref{fig:eva}), the proposed scheme reduces the BER by more than 2 orders of magnitude thanks to stage 2 iterative detections, whereas the former suffers from band approximation induced ISI. In comparison with OFDM with the same band approximation on channel matrix and FD equalization, AFDM manifests significant performance advantage due to better diversity exploitation.

\begin{figure}
\centering
    \includegraphics[width=3.0in]{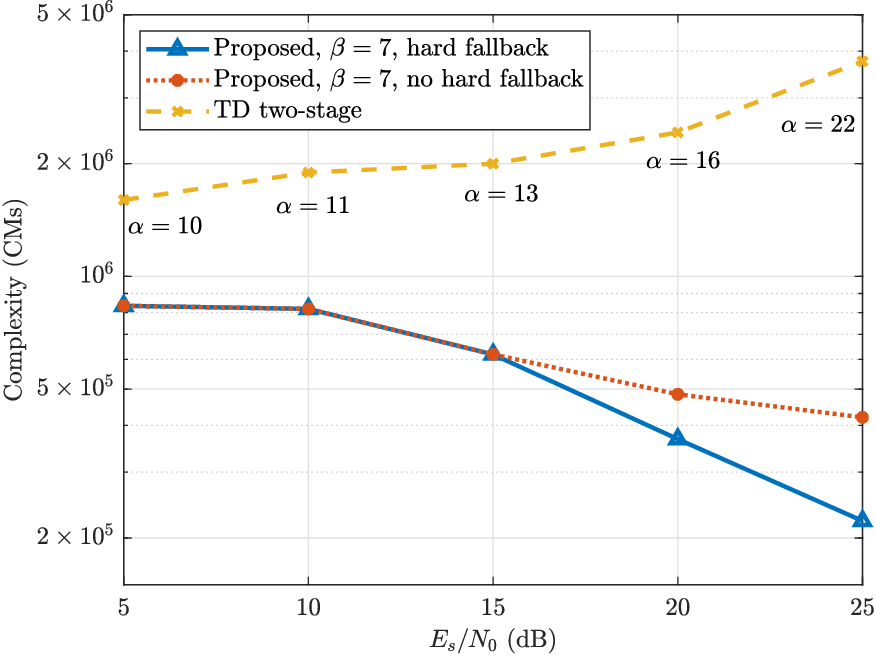}
    \vspace{-2mm}
\caption{Required computational complexity of different equalizers to achieve the same BER with the proposed FD equalizer (\(\beta=7\)) in Fig.~\ref{fig:eva} at different \(E_s/N_0\).}
\label{fig:complexity_ber}
\end{figure}

We also provide a sophisticated comparison between TD and FD equalization using the same two-stage detection framework. From a controlled complexity perspective, we first apply a band approximation with \(\alpha=7\) to the TD channel matrix (demonstrated in Fig.~\ref{fig:ch_mtx}), and employ the same two-stage equalization except operating in TD, so that the equalization has about the same computational complexity as the proposed FD-based scheme. However, as shown in Fig.~\ref{fig:eva}, the TD equalizer reaches a significantly high error floor, since the band approximation is ineffective due to a large actual bandwidth of the TD channel matrix, as per Section~\ref{sec:3}. On the other hand, from a common target BER perspective, we increase \(\alpha\) for the band approximation TD channel matrix until the equalizer surpasses the performance of the proposed FD equalizer with \(\beta=7\) at different \(E_s/N_0\). The \(\alpha\) values and corresponding computational complexity (averaged between random realizations due to varying iteration numbers) are shown in Fig.~\ref{fig:complexity_ber}. As can be observed, an increasingly larger \(\alpha\) is demanded for the TD equalizer to keep up with BER performance of the proposed FD equalizer as \(E_s/N_0\) increases. While the required number of iterations decreases at higher \(E_s/N_0\), the increase in \(\alpha\) still causes the complexity for TD equalizer to increase. In comparison, the proposed FD equalizer reduces the complexity by more than one order of magnitude, with a deterministically small value of \(\beta\). Note that DAFT domain equalization is omitted due to space limitations, as it has been shown to be inferior to TD in both performance and complexity~\cite{TZP_AFDM}, which is also reflected by the channel-matrix structures as per Fig.~\ref{fig:ch_mtx}. 

Finally, with the introduction of hard-decision  fallback activated for \(\varepsilon<0.1\), Fig.~\ref{fig:eva} shows that it does not have noticeable impact on the error performance of the proposed scheme, while Fig.~\ref{fig:complexity_ber} demonstrates its complexity reduction by approximately a half at high \(E_s/N_0\) comparing to non-fallback scheme.

\section{Conclusion}
In this paper, we investigated frequency domain equalization for AFDM under general physical channels. We derived the IO relations of the considered filtered-AFDM waveform across different domains, and showed that under off-grid delay and Doppler shifts, the DAFT domain equivalent channel exhibits widespread ISI following the combined effect of the composition pulse and Dirichlet functions. Through a representative example, we further demonstrated that the frequency domain can be more favorable for low-complexity equalization in wideband systems, owing to its more compact channel representation compared to the time and DAFT domains. Motivated by this observation, we developed a two-stage FD equalizer, where the first stage employs block Cholesky factorization to obtain an accurate initial estimate, and the second stage performs cross-domain equalization, with a hard-decision fallback mechanism to reduce complexity. Finally, we demonstrated that the proposed scheme achieves error performance close to full-block LMMSE at substantially lower complexity. Moreover, it outperforms time domain equalization at the same complexity level, while requiring lower computational cost to achieve a given BER target.
\bibliographystyle{IEEEtran}
\bibliography{ref}

@STRING{J_IEEE_TWC       = "{IEEE} Trans. Wireless Commun."}

@STRING{J_IEEE_TCOM        = "{IEEE} Trans. Commun."}

@String{C_IEEE_ICC               = "Proc. {IEEE} Int. Conf. Commun. (ICC)" }

@ARTICLE{oddm,
  author={Lin, Hai and Yuan, Jinhong},
  journal=J_IEEE_TWC,
  title={Orthogonal Delay-{D}oppler Division Multiplexing Modulation},
  year={Dec. 2022},
  volume={21},
  number={12},
  pages={11024-11037},
  doi={10.1109/TWC.2022.3188776}}

@inproceedings{Hadaniwcnc17,
author = {R. Hadani and others},
title = {Orthogonal time frequency space modulation},
booktitle = {Proc. of IEEE WCNC},
year = {2017}
}

@ARTICLE{otfs_xdomain,
  author={Li, Shuangyang and others},
  journal=J_IEEE_TWC, 
  title={Cross Domain Iterative Detection for Orthogonal Time Frequency Space Modulation}, 
  year={2022},
  volume={21},
  number={4},
  pages={2227-2242},
  doi={10.1109/TWC.2021.3110125}}

@ARTICLE{bello,
  author={Bello, P.},
  journal={IEEE Trans. Commun. Syst.}, 
  title={Characterization of Randomly Time-Variant Linear Channels}, 
  year={1963},
  volume={11},
  number={4},
  pages={360-393},
  doi={10.1109/TCOM.1963.1088793}}

@ARTICLE{oddm_io,
  author={Tong, Jun and others},
  journal=J_IEEE_TCOM, 
  title={Orthogonal Delay-Doppler Division Multiplexing ({ODDM}) Over General Physical Channels}, 
  year={2024},
  volume={},
  number={},
  pages={1-1},
  doi={10.1109/TCOMM.2024.3416892}}

@ARTICLE{adaptive_FD,
  author={Zhang, Hongyang and others},
  journal=J_IEEE_TWC, 
  title={Adaptive Transmission With Frequency-Domain Precoding and Linear Equalization Over Fast Fading Channels}, 
  year={2021},
  volume={20},
  number={11},
  pages={7420-7430},
  doi={10.1109/TWC.2021.3083652}}

@ARTICLE{AFDM_MP,
    author={Wu, L and others},
    title={A Message Passing Detection based Affine Frequency Division Multiplexing Communication System},
    year={2023},
url={https://doi.org/10.48550/arXiv.2308.01802}}

@ARTICLE{AFDM,
  author={Bemani, Ali and Ksairi, Nassar and Kountouris, Marios},
  journal=J_IEEE_TWC, 
  title={Affine Frequency Division Multiplexing for Next Generation Wireless Communications}, 
  year={2023},
  volume={22},
  number={11},
  pages={8214-8229},
  doi={10.1109/TWC.2023.3260906}}

@INPROCEEDINGS{AFDM_LDL,
  author={Bemani, Ali and Ksairi, Nassar and Kountouris, Marios},
  booktitle={ICASSP}, 
  title={Low Complexity Equalization for Afdm In Doubly Dispersive Channels}, 
  year={2022},
  volume={},
  number={},
  pages={5273-5277},
  doi={10.1109/ICASSP43922.2022.9746329}}

@book{matrix_comp,
  title        = {Matrix Computations},
  author       = {Golub, Gene H. and Van Loan, Charles F.},
  volume       = {3},
  year         = {2012},
  publisher    = {Johns Hopkins University Press},
  address      = {Baltimore, MD, USA}
}

@INPROCEEDINGS{adpt_2,
  author={Shen, Cheng and Shafie, Akram and Yuan, Jinhong},
  booktitle=C_IEEE_ICC, 
  title={Channel-Dependent Adaptive Time/Frequency Domain Detection for {ODDM}}, 
  year={2025},
  volume={},
  number={},
  pages={983-988},
  doi={10.1109/ICC52391.2025.11161823}}

@ARTICLE{TZP_AFDM,
  author={Shen, Cheng and Yuan, Jinhong and Tong, Jun},
  journal=J_IEEE_TWC, 
  title={Time-Domain Zero-Padding ({TZP}) {AFDM} With Two-Stage Iterative MMSE Detection}, 
  year={2026},
  volume={25},
  number={},
  pages={6255-6269},
  doi={10.1109/TWC.2025.3624211}}

@INPROCEEDINGS{fOFDM,
  author={Zhang, Xi and others},
  booktitle={IEEE Globecom}, 
  title={Filtered-{OFDM} - Enabler for Flexible Waveform in the 5th Generation Cellular Networks}, 
  year={2015},
  volume={},
  number={},
  pages={1-6},
  doi={10.1109/GLOCOM.2015.7417854}}

\end{document}